\documentclass[12pt]{elsart}
\usepackage{epsfig,bm,amsfonts,amssymb,amsmath}

\begin{document}
\runauthor{Carmona, Cort\'es, Gamboa, M\'endez}
\begin{frontmatter}
\title{Noncommutativity in Field Space and Lorentz Invariance Violation}
\author{J.M. Carmona and J.L. Cort\'es}
\address{Departamento de F\'{\i}sica Te\'orica, Universidad de Zaragoza,
50009 Zaragoza, Spain}
\ead{jcarmona,cortes@posta.unizar.es}
\author{J. Gamboa and F. M\'endez}
\address{Departamento de F\'{\i}sica, Universidad de Santiago de Chile,
Casilla 307, Santiago 2, Chile}
\ead{gamboa,fmendez@lauca.usach.cl}

\begin{abstract}
The connection between Lorentz invariance violation and noncommutativity
of fields in a quantum field theory is investigated.
A new dispersion relation for a free field theory with just one
additional noncommutative parameter is obtained.
While values for the noncommutative scale much larger than
$10^{-20} \,\mathrm{eV}^{-1}$ are ruled out by the present experimental
status, cosmic ray physics would be compatible with and sensible to
a noncommutativity arising from quantum gravity effects.
We explore the matter-antimatter asymmetry which is naturally present
in this framework.
\end{abstract}

\begin{keyword}
Noncommutativity \sep Lorentz invariance violation \sep Field theory
\sep Cosmic rays
\PACS 04.60.-m \sep 11.30.Cp \sep 98.70.Sa
\end{keyword}
\end{frontmatter}

Recently several groups have argued that quantum gravity relics
could be seen from dispersion relations violating Lorentz
invariance~\cite{1}. In order to explain these results, they used
dispersion relations coming from loop quantum gravity or other arguments
based on effective field theories. Lorentz invariance is then seen
as a good low-energy symmetry which may be violated at very high
energies~\cite{kos}. This violation is usually compatible with translational
and rotational invariance in a ``preferred frame''.
Since our low-energy theories are relativistic quantum field 
theories (QFT's),
it is interesting to explore possible extensions of the QFT
framework which could produce departures from exact Lorentz invariance.

The purpose of this letter is to show how the assumption of
noncommutativity (NC) in the field space of a QFT
produces Lorentz-violating dispersion relations. Moreover, we will
show that cosmic ray physics is sensitive to a NC scale as low
as the Planck length. In particular a consequence of NC
at this scale would be the absence of the GZK cutoff~\cite{gzk} in the
cosmic ray spectrum. NC will moreover be a possible source
of an asymmetry between matter and antimatter, with physical processes
distinguishing between them.

Let us firstly consider the theory of a complex scalar free field
on a nonconmutative space (i.e. $[x_i,x_j]=i\theta_{ij}$) described
by the following action
\begin{equation}
\mathcal{S} = \int d^4 x \, \left[\partial_\mu  \phi^* \star\partial^\mu \phi -
m^2 \phi^* \star \phi\right]. \label{3}
\end{equation}

Owing to the properties of the Moyal product
\begin{equation}
({\bm A}\star {\bm B})({\bm x}) =
\lim_{{\bm x_1},{\bm x_2} \rightarrow {\bm x}} e^{\frac{i}{2}\theta^{ij}\partial^{(1)}_i
\partial^{(2)}_j}{\bm A}({\bm x}_1) {\bm B}({\bm x}_2), \label{4}
\end{equation}
the action~(\ref{3}) is equivalent to the commutative one and, as a
consequence, one concludes that free field theory
cannot be modified by spacetime NC~\cite{szabo}.

There is however another way to introduce NC in a
QFT. We will illustrate this with an example
in quantum mechanics. Let us consider the Hamiltonian of an harmonic
oscillator in two dimensions,
\begin{equation}
H=\frac{1}{2}\,(p_1^2+p_2^2)+\frac{1}{2}\,(q_1^2+q_2^2),
\label{oscilador}
\end{equation}
where the $(q_i,p_i)$ satisfy the canonical commutation relations
\begin{subequations}
\begin{eqnarray}
\left[p_i ,p_j \right] &=& 0, \label{co3}
\\
\left[q_i ,q_j \right] &=& 0, \label{co1}
\\
\left[q_i,p_j \right] &=& i \delta_{ij}. \label{co2}
\end{eqnarray}
\end{subequations}
In the noncommutative space, one deforms the commutator (\ref{co1}) as follows
\begin{equation}
\left[q_i ,q_j \right] = i \epsilon_{ij} \theta, \label{co4}
\end{equation}
where $\theta$ is a measure of the spatial NC. However, 
since Eq.~(\ref{oscilador}) is invariant under the symmetry $q_i
\leftrightarrow p_i$, then this is equivalent to a deformation of the
commutator~(\ref{co3}):
\begin{equation}
\left[p_i,p_j\right] = i \epsilon_{ij} \theta, \label{co5}
\end{equation}
and, as a consequence, the commutative and noncommutative two dimensional
oscillators are very different systems~\cite{nair}.

Note that Eq.~(\ref{co4}) is a relation at the level of the degrees
of freedom of the system. Analogously, the introduction of NC in a QFT 
at the level of the fields, which is
different from the NC in spatial coordinates,
will produce nontrivial modifications of the QFT
framework already at the level of a free theory.

Let us then come back to QFT and consider the Hamiltonian of the free
complex bosonic field theory
\begin{equation}
\mathcal{H}=\frac{1}{2}\,\Pi_i^2+\frac{1}{2}\bm{\nabla}\Phi_i\cdot
\bm{\nabla}\Phi_i+\frac{m^2}{2}\Phi_i^2
\label{Hnuevo}
\end{equation}
where a sum over the two field components (the real and imaginary parts of
the original complex field) is assumed. In the Shr\"odinger 
representation~\cite{hattfield},
for fields satisfying the usual canonical commutation relations, 
this Hamiltonian becomes the operator
\begin{equation}
\hat{H}_C= \frac{1}2\int d{\bm x}\,
\left[- \frac{\delta^2}{\delta \phi _i(x)^2}+
(\bm{\nabla} \phi_i)^2 + m^2 \phi_i^2 \right], \label{11}
\end{equation}
acting on functionals of classical fields $\phi$, which are the eigenvalues
of the $\Phi$ field operators.

We may introduce now a NC in field theory.
The simplest option is to deform the commutator of fields in analogy with the
deformation of the commutator of the coordinates, Eq.~(\ref{co4}),
in the previous quantum mechanical example. We might want to preserve
the locality in the new set of canonical commutation relations, which
become
\begin{subequations}
\begin{eqnarray}
[\Pi_i(x),\Pi_j(x')] &=& 0, \label{121} \\
\mbox{[}\Phi_i(x),\Phi_j(x')\mbox{]}&=& i\,
{\bar \theta} \,\epsilon_{ij} \,\delta({\bm x},{\bm x'}),\label{12} \\
\mbox{[}\Phi_i(x'),\Pi_j(x)\mbox{]}&=&i\,\delta_{ij}\,\delta({\bm x},{\bm x'}).
\end{eqnarray} \label{122}
\end{subequations}
The NC parameter $\bar\theta$ has the dimension of
$\sqrt{\theta}$, where $\theta$ is the usual parameter of NC
in quantum mechanics.

The Moyal product Eq.~(\ref{4}) allows to map the study of noncommutative
field theories into that of ordinary field theories where the ordinary 
product is replaced by the star product~\cite{connes,sw}. The fact that the 
noncommutativity of the base manifold can be bypassed with the help of the 
star operation may be also used to define a noncommutative quantum 
mechanics~\cite{mezincescu}. One can do the same trick here. We must propose
a new Moyal product between functionals consistent with the commutation
relations~(\ref{122}). Defining
\begin{equation}
\Psi_1[\phi]\star \Psi_2[\phi]=\lim_{\eta,\, \xi \rightarrow \phi}
e^{\frac{i}{2}{\bar \theta} \epsilon_{ij}
\int d{\bm x}\frac{\delta}{\delta \eta_{i}(x)}
\frac{\delta}{\delta\xi_{j}(x)} } \Psi_1[\eta]\Psi_2[\xi], \label{13}
\end{equation}
we verify straightforwardly
\begin{equation}
[\phi_i(x),\phi_j(x')]_{\star}\equiv
\phi_i(x)\star\phi_j(x')-\phi_j(x')\star\phi_i(x)
=i\,{\bar \theta} \epsilon_{ij} \,\delta({\bm x},{\bm x'}),  \label{14}
\end{equation}
and the standard properties of the Moyal product hold.
One should note however that this star product is completely different from
that of Eq.~(\ref{4}).

Given Eq.~(\ref{13}), the functional Schr\"odinger equation
in the field configuration space becomes
\begin{equation}
\frac{1}2\int d{\bm x} \bigg[- \frac{\delta^2}{\delta \phi_i (x)^2}+
(\bm{\nabla} \phi_i)^2 + m^2 \phi_i^2\bigg ] \star
\Psi[\phi_j] = E \Psi[\phi_j],
\label{15}
\end{equation}
where one should understand that the functional derivatives remain unchanged
under the star operation.

The noncommutativity introduced by the star product in Eq.~(\ref{15}) 
is equivalent to replacing the Hamiltonian operator~(\ref{11}) by 
a new ``noncommutative'' Hamiltonian
\begin{eqnarray}
{\hat H}_{NC}&=&{\hat H}_C + \frac{1}2\int d{\bm x}\bigg[ i {\bar \theta}
\epsilon_{ij}\bigg(m^2 \phi_i \frac{\delta}{\delta\phi_j} + \bm{\nabla}
\phi_i \bm{\nabla}\bigg[\frac{\delta}{\delta\phi_j}\bigg]\bigg) \nonumber
\\
&-&\frac{{\bar \theta}^2 }{4}\, \epsilon_{ij}\epsilon_{ik}\bigg( \bm{\nabla}
\bigg[\frac{\delta }{\delta\phi_{j}(x)}\bigg]\bm{\nabla}
\bigg[\frac{\delta}{\delta\phi_{k}(x)}\bigg] 
+m^2\frac{\delta^2}{\delta
\phi_{j}(x)\delta\phi_k(x)}\bigg)\bigg], \label{18}
\end{eqnarray}
acting on the same space of functionals of classical fields. 

We have mapped the original theory given by the Hamiltonian~(\ref{Hnuevo}) 
in terms of noncommutative fields~(\ref{122}) into a theory of ordinary
fields having $\hat{H}_{NC}$ as the Hamiltonian in its Schr\"odinger 
representation. This has some analogy with the Seiberg-Witten map~\cite{sw}
in Yang-Mills theories, which associates to every noncommutative gauge
theory an ordinary gauge theory with a modified Hamiltonian.

Assuming that the $\bar\theta\to 0$ limit is not 
singular\footnote{See ``Note added'' at the end of this letter.}, the theory 
described by the Hamiltonian~(\ref{18}) will be a theory of free particles.
Using translational invariance,
one can derive their dispersion relation through the common eigenvalues of
$\hat{H}_{NC}$ and the momentum operator. The correspondence between classical
and quantum theories allows to determine these eigenvalues easily by 
solving the evolution equation of the classical Hamiltonian 
\begin{eqnarray}
H_{NC}&=&\frac{1}2\int d^3{\bm x}\bigg[ \pi^2
- {\bar \theta} \epsilon_{ij}\bigg(\bm{\nabla} \phi_i \bm{\nabla}\pi_j
+  m^2 \phi_i\pi_j\bigg) \nonumber \\
&+& (\bm{\nabla} \phi)^2 + m^2 \phi^2
+ \frac{{\bar \theta}^2}4\bigg((\bm{\nabla}\pi)^2+
m^2 \pi^2 \bigg)\bigg], \label{19}
\end{eqnarray}
with a plane wave ansatz
\begin{equation}
\phi_i (\bm{x},t)=A_i e^{i\,E\, t - i \bm{k}\cdot\bm{x}}\, , \quad
\pi_i (\bm{x},t)=B_ie^{i\,E\, t - i \bm{k}\cdot\bm{x}}\, .  \label{li}
\end{equation}
Defining $k\equiv|\bm{k}|$, we obtain
\begin{eqnarray}
0&=& B_1 +\frac{\bar \theta}2(m^2 +k^2)A_2 +\frac{\bar \theta^2}4\,
(m^2 +k^2)B_1 - i\,E\, A_1,\nonumber
\\
0&=& B_2 -\frac{\bar \theta}2(m^2 +k^2)A_1+\frac{{\bar \theta}^2}4\,
(m^2+k^2) \,B_2 -i\,E\, A_2,\nonumber
\\
0&=& (m^2+k^2)A_1- \frac{\bar \theta}2(m^2+k^2)B_2+i \,E\,B_1,\nonumber
\\
0&=& (m^2+k^2)A_2+\frac{\bar \theta}2(m^2+k^2)B_1+i \,E\,B_2.
\label{20}
\end{eqnarray}
The energies $E(k)$ obtained as a solution of this linear system of equations
are exactly the eigenvalues of the Hamiltonian $\hat H_{NC}$, with $\bm{k}$
being the corresponding eigenvalues of the momentum operator. This can be
easily proved from the correspondence between classical fields and 
matrix elements of the field operator.

Nontrivial solutions of the previous set of equations are obtained if
and only if the principal determinant vanishes. Note that under the
exchange $A_1\leftrightarrow A_2, B_1\leftrightarrow B_2$, the coefficient
matrix in Eq.~(\ref{20}) remains identical if we change $\bar\theta$ by
$-\bar\theta$. Therefore if its determinant vanishes for a certain
$\bar\theta$, it will also do so for $-\bar\theta$.
This is reflected in the double solution indicated by the $\pm$ sign
(we take from now on $\bar\theta>0$) in the following dispersion relation
obtained from the vanishing condition of the determinant:
\begin{equation}
E^2_\pm=k^2 + m^2 +
\frac{{\left( k^2 + m^2 \right) }^2\,{\bar \theta }^2}{2} 
\pm \frac{{\left( k^2 + m^2 \right) }^{\frac{3}{2}}\,{\bar \theta} \,
{\sqrt{4 + \left( k^2 + m^2 \right) \,{\bar \theta}^2}}}{2}\, .
\label{21}
\end{equation}
This dispersion relation is not Lorentz invariant, which could have
been anticipated either from the commutation relations~(\ref{122}),
which are no longer covariant owing to the different Lorentz transformation
laws for the field and the momentum,
or, alternatively, from the Lorentz noninvariant terms in the
Hamiltonian~(\ref{19}).

Since the interchange $\phi_1\leftrightarrow\phi_2$ corresponds
(appart from a global $i$ factor)
to the exchange $\phi\leftrightarrow\phi^*$, we conclude that the term
proportional to $\bar\theta$ in Eq.~(\ref{21}) will be of opposite sign
for the particle and antiparticle described by the complex field $\phi$.
This matter-antimatter asymmetry which comes out from the NC
will have important consequences as we will explore later.

Let us simplify Eq.~(\ref{21}) a little bit. Defining
$E_{LI}\equiv\sqrt{k^2+m^2}$, we see that $E_{\pm}^2/E_{LI}^2$ is a function
of the variable $a\equiv E_{LI}\bar\theta$. In fact
\begin{equation}
\frac{E_{\pm}^2}{E_{LI}^2} = 1 + \frac{a^2}{2}\pm\frac{a\sqrt{4+a^2}}{2}.
\label{fdef1}
\end{equation}

Lorentz invariance is a very good low-energy approximation, well tested
at least up to the TeV scale. This puts an upper
bound on the NC parameter,
$\bar\theta\lesssim 10^{-12} \,\mathrm{eV}^{-1}$.
Small deviations mean $a\ll 1$. Up to order $a^3$, we therefore get
\begin{equation}
E^2_{\pm} = k^2 + m^2 \pm (k^2+m^2)^{3/2}\,\bar\theta+
(k^2+m^2)^2\,\frac{\bar\theta^2}{2}\,.
\label{newdisp}
\end{equation}

In the ultrarelativistic limit, $k\gg m$, keeping only each of the first terms
in the expansion in both powers of $m^2$ and $\bar\theta$ of
Eq.~(\ref{newdisp}), we get
\begin{equation}
E_\pm = k + \frac{m^2}{2k}\pm\frac{1}{2}\,\bar\theta k^2.
\label{URlimit}
\end{equation}
In the nonrelativistic limit, $k\ll m$, then expanding Eq.~(\ref{newdisp})
in powers of $k^2$ and $\bar\theta$, we obtain to first order
\begin{equation}
E_{\pm} = m + \frac{k^2}{2m} \pm \frac{1}{2}\,\bar\theta\, m^2.
\label{NURlimit}
\end{equation}

In contrast with a general parametrization of a Lorentz-violating dispersion
relation involving many undetermined coefficients~\cite{cg}, the
phenomenological analysis of a Lorentz invariance violation (LIV) produced
by NC in field space is more economical because the only
coefficient present in Eq.~(\ref{21}) is the NC parameter
${\bar \theta} $ and, therefore, one could search for an appropriate
experiment in order to extract some limits on it.

The LIV's induced by the NC are
a high-energy effect (of scale $\bar\theta^{-1}$). They could therefore
be observed in high-energy experiments or in low-energy experiments of
very high precision. Complete analyses of the noncommutative effects to
low-energy experiments would be quite involved, and in particular would
require to go beyond free theory. But in order to get an estimate of 
the sensitivity to $\bar\theta$ at low-energies it is enough to concentrate
on a simple experiment, such as the tritium beta-decay, and study the
modifications induced by the noncommutative dispersion relation 
Eq.~(\ref{21}). The tritium beta decay is a low-energy
experiment~\cite{tri}, with scales the electron mass and $Q$, the
total energy available for the process. However,
its high precision makes it sensitive to tiny deviations induced by
a very small non-zero mass from the $E=k$ relation for a
massless neutrino. What about the small deviation from the relativistic
dispersion relation coming from the NC? Since for a neutrino
$k\gg m$, the energy-momentum relation is Eq.~(\ref{URlimit}).
The mass term gets important at the end of the beta-decay spectrum
(small $k$ for the neutrino), while the NC term is important
at the beginning of the spectrum. Both corrections are comparable at a neutrino
momentum given by
\begin{equation}
k\sim \left(\frac{m}{\,\mathrm{eV}}\right)^{2/3}
\left[\frac{(10^{13}\,\mathrm{eV})^{-1}}{\bar\theta}\right]^{1/3}
(20 \, \mathrm{keV}).
\end{equation}
Since 20 keV is the order of magnitude of the maximum momentum for the
neutrino in the tritium beta-decay, the possibility to detect a parameter
of NC $\bar\theta\sim (10^{13}\,\mathrm{eV})^{-1}$ 
using data from the beginning of the electron energy spectrum 
is comparable to the possibility to detect a neutrino mass of order the eV 
using the experimental data from the tail of the spectrum.
In fact,  this $\bar\theta$ would require a precision in the
determination of the energy of the neutrino
$\delta E_\nu\sim 10^{-4} - 10^{-5} \,\mathrm{eV}$, far from the experimental
possibilities.

An alternative would be to consider the dispersion relation for the
electron. Since in the tritium beta-decay the electron is nonrelativistic,
its energy-momentum relation is Eq.~(\ref{NURlimit}).
Detection of $\bar\theta\sim (10^{13}\,\mathrm{eV})^{-1}$ here requires
a determination of the energy of the electron with a
$10^{-1}-10^{-2} \,\mathrm{eV}$ precision. But there is an extra difficulty
here: the modification in the energy is a constant (acting as an additional
``effective'' mass) that cannot be distinguished
from a change in the $Q$ factor, whose determination certainly contains
errors greater than the eV.

On the other hand, the mass of the electron is known with a precision of
$2\times 10^{-2}$ eV. So a $\bar\theta \sim (10^{13}\,\mathrm{eV})^{-1}$
would be detectable from the experiments used to measure the electron mass
if they were able to separate the effective mass coming from the
NC. If this were so also in the case of the proton, even a
$\bar\theta\sim (10^{16}\,\mathrm{eV})^{-1}$ would be detectable (since
the ``effective'' mass, $\approx m_p^2\bar\theta$, is larger in this case).
But to see this,  one should go beyond the free theory in the discussion of
noncommutative quantum fields. We turn instead to the possibility of
direct exploration of scales $1/\bar\theta$ offered by the physics of
high-energy cosmic rays, which is known to change drastically by violation of
Lorentz symmetry~\cite{cg,abcc}. We will see how this physics is sensitive 
to much lower values of the NC parameter $\bar\theta$.

For high energy cosmic rays ($E_{LI} \sim k \geq 10^{16} \,\mathrm{eV}$) and
a NC parameter
$\bar\theta\sim (10^{13}\,\mathrm{eV})^{-1}$, we are no longer in the limit
$a\ll 1$; instead we have $k\bar\theta\gg 1$. The energy-momentum relation
for this case can be easily obtained from Eq.~(\ref{fdef1}):
\begin{equation}
E_+ \approx k^2\bar\theta\, , \quad
E_-^2 \approx \bar\theta^{-2}-2/(k^2\bar\theta^4).
\end{equation}
Experimental observation of cosmic rays up to energies
$E\sim 10^{20}\,\mathrm{eV}$ rules out both solutions \cite{halzen}: with the $E_+$ relation,
any disintegration of one particle into two particles allowed by the
conservation laws of the strong interaction (e.g. $p\to n + \pi^+$) is also
allowed by energy conservation, whatever their masses are. This is
an effective mechanism of energy loss for primary cosmic rays. A particle
with the $E_-$ relation would have a maximum energy of $1/\bar\theta$.
Therefore the parameter of NC cannot be much higher than
$(10^{20}\,\mathrm{eV})^{-1}$. In particular the values
$\bar\theta\sim (10^{13}\,\mathrm{eV})^{-1}$ are completely excluded.

An interesting possibility is that the NC would
be a quantum gravity effect, so that  $\bar\theta^{-1}$ would naturally
be of order the Planck scale, $M_P\sim 10^{28}\,\mathrm{eV}$. In general, for
$\bar\theta\ll (10^{20}\,\mathrm{eV})^{-1}$, we can use the approximation
$a\ll 1$ in the physics of high-energy cosmic rays and the energy-momentum
relation given by Eq.~(\ref{URlimit}). Let us con\-sider the kinematics
of the disintegration of a particle of mass $m$ into two particles
$m_1$ and $m_2$ of momenta $\bm{p}_1$
and $\bm{p}_2$ with relative angle $\phi$. Energy conservation leads to
\begin{equation}
\cos\phi\approx 1
+ \left(\frac{m_1^2}{2p_1}+\frac{m_2^2}{2p_2}-\frac{m^2}{2(p_1+p_2)}\right)
\left(\frac{1}{p_1}+\frac{1}{p_2}\right)
\mp \bar\theta\,(p_1+p_2).
\label{condicion}
\end{equation}
Without the NC contribution, $\cos\phi>1$ if $m_1+m_2>m$
and the disintegration is forbidden, while it is kinematically allowed
if $m_1+m_2<m$. A positive correction to the energy in Eq.~(\ref{URlimit})
coming from the NC enlarges the range of masses for which the
disintegration is allowed, while the negative correction produces
the inverse effect: kinematically allowed disintegrations in the
relativistic invariant theory are no longer allowed above a certain value
of the momentum.

Taking masses of the order of the GeV, the momentum scale
$p_{nc}$ above which: either disintegration is an effective mechanism for
energy loss in $E_+$ case, or kinematically allowed disintegrations
in the Lorentz invariant theory are forbidden because of the
NC in the $E_-$ case, is
\begin{equation}
p_{nc}\sim \left[\frac{(10^{28}\,\mathrm{eV})^{-1}}{\bar\theta}\right]^{1/3}
2\times 10^{15} \,\mathrm{eV}.
\end{equation}
Again, the existence of cosmic rays of energies as high as
$E\sim 10^{20}\,\mathrm{eV}$ discards the $E_+$ energy-momentum relation
for them [at least for values $\bar\theta\geq (10^{43}\,\mathrm{eV})^{-1}$],
while the $E_-$ relation would still be compatible with the observation of
such energetic cosmic rays. In this last case,  we could consider the
energy loss mechanism coming from the interaction with the microwave
background of photons, which in standard Lorentz invariant kinematics
produces the GZK cutoff~\cite{gzk} at $10^{19}\,\mathrm{eV}$. Incorporating
the energy of the photon $\omega\sim 10^{-4}\,\mathrm{eV}$ to the
energy balance Eq.~(\ref{condicion}), the reaction
$p + \gamma \to n + \pi^+$ is kinematically allowed with the $E_-$
relation coming from NC if
\begin{equation}
\left[\omega+\frac{m^2}{2(p_1+p_2)}-\frac{m_1^2}{2p_1}
-\frac{m_2^2}{2p_2}\right]\geq \bar\theta \,p_1 p_2.
\label{gzk}
\end{equation}
Without NC, the right hand side (rhs) of Eq.~(\ref{gzk}) is zero
and the inequality is satisfied above a certain value of the momentum
$p_1+p_2$ (the standard GZK cutoff). However, for values of the NC
parameter $\bar\theta \geq (10^{43}\,\mathrm{eV})^{-1}$, Eq.~(\ref{gzk})
is no longer satisfied for any momentum: the GZK cutoff disappears.

In summary, for the energy-momentum relation Eq.~(\ref{21}) which
comes out from noncommutative quantum fields [Eq.~(\ref{122})],
only the $E_-$ solution for particles would be
compatible with the observation of very high-energy cosmic rays.
Moreover, the NC washes out the standard GZK cutoff,
and is therefore a possible explanation of why this cutoff is not
observed in the cosmic ray spectrum. Recalling the comments
following Eq.~(\ref{20}), the $E_+$ relation should be assigned to
antiparticles. According to our previous discussion, they would
present a very efficient energy loss mechanism at high energies
(disintegration by strong processes) and could not propagate
through very large distances. These conclusions apply to very tiny
values of the NC parameter [the limitation is just
$\bar\theta\geq (10^{43}\,\mathrm{eV})^{-1})$], including the
suggestive scenario in which NC arises from
quantum gravity effects ($\bar\theta^{-1}\sim M_P$).

We should note that for certain particles the assignation
of the $E_-$ or $E_+$ energy-momentum relations is still ambiguous.
For example, the distinction between matter and antimatter for the
$\pi^+$ and the $\pi^-$ is not clear.
We used the $E_-$ relation for the $\pi^+$ in Eq.~(\ref{gzk}),
but we could have equally chosen the $E_+$ relation for it.
Moreover, the $\pi^0$ is described by a real field, so that we cannot
make it noncommutative (according to Eq.~(\ref{12}) we need two-component
fields for the introduction of noncommutativity). Since the $\pi^0$,
together with the $\pi^+$ and the $\pi^-$ forms an
SU(2) triplet, one could argue the absence of noncommutative corrections
for the three particles based on symmetry arguments. However, in this
case the ambiguity does not affect the conclusion of the disappearance
of the GZK cutoff. Taking the $E_+$ relation for the $\pi^+$ modifies
the rhs of Eq.~(\ref{gzk}) by the factor $(1+p_2/p_1)$.
The factor is $(1+p_2/2p_1)$ if the $\pi^+$ does not present any
$\bar\theta$ correction. In both cases the corresponding inequality
is even more difficult to satisfy than Eq.~(\ref{gzk}).
On the other hand, since we are considering different
dispersion relations for particles
and antiparticles, a very stringent bound
($\bar\theta \lesssim 10^{-27} \mathrm{eV}^{-1}$) is given by
kinematic CPT violation in neutral kaons~\cite{kaon}. If however
one extends the previous symmetry arguments for pions to SU(3) then there
are no bounds to $\bar\theta$ coming from kaon physics.

Finally, let us comment that QFT formulated in a
noncommutative space also produces violations of relativistic invariance
if one goes beyond free theory. This is a ``dynamic''
LIV, in the sense that it is produced only
in the presence of an interaction. Existing experiments bound the
NC energy scale to $10^{13}$ eV~\cite{kost}.
In contrast, NC
in field space produces a ``kinematic'' violation, which allows the
identification of corrections to the relativistic dispersion relation
containing a single parameter $\bar\theta$. In this case high-energy cosmic
rays put a much more stringent bound on the NC energy scale
$\bar\theta^{-1}$.

We would like to thank Prof M. Loewe for discussions. 
This work has been partially supported by the
grants 1010596, 7010596 and 3000005 from Fondecyt-Chile,
by M.AA.EE./AECI and 
by MCYT (Spain), grants FPA2000-1252 and FPA2001-1813.

\textbf{Note added:} After this work was submitted for publication,
we proved the validity of the hypothesis that the Hamiltonian~(\ref{18})
describes a theory of free particles by defining 
an explicit quantization in Fock space \cite{QTNCF}. This reformulation
allowed us to go beyond the simplest option of noncommutativity by
including a nontrivial commutator for the momenta [a deformation of
the relation~(\ref{121})].


\begin{thebibliography}{}
\bibitem{1} G. Amelino-Camelia, J. Ellis, N.E. Mavromatos, D.V. Nanopoulos
and S. Sarkar, Nature {\bf393}, 763 (1998);
R. Gambini and J. Pullin, Phys. Rev. {\bf 59}, 124021 (1999);
J. Alfaro, H. A. Morales-T\'ecotl and L. F. Urrutia,
Phys. Rev. Lett. {\bf 84}, 2318 (2000);
J. Alfaro, H. A. Morales-T\'ecotl and L. F. Urrutia, 
Phys. Rev. D {\bf 65}, 103509 (2002);
J. Alfaro and G. Palma, Phys. Rev. D {\bf 65}, 103516 (2002);
T. Jacobson, S. Liberati, D. Mattingly, hep-ph/0112207.
\bibitem{kos} The problem of Lorentz violation invariance has been 
extensively studied in the last years by
Kosteleck\'y and collaborators. Some references are:  
D. Colladay and V.A. Kosteleck\'y,  Phys. Lett. {\bf B511}, 209 (2001); 
V.A. Kosteleck\'y, R. Lehnert, Phys. Rev. D {\bf 63}, 065008 (2001); 
R. Bluhm and  V.A. Kosteleck\'y, Phys. Rev. Lett. {\bf 84}, 1381 (2000);
V.A. Kosteleck\'y and  Charles D. Lane, Phys. Rev. D {\bf 60}, 116010 (1999);
R. Jackiw and V.A. Kosteleck\'y, Phys. Rev. Lett. {\bf 82}, 3572 (1999); 
D. Colladay, V.A.  Kosteleck\'y, Phys. Rev. D {\bf 58}, 116002 (1998).
\bibitem{gzk} K. Greisen, Phys. Rev. Lett. {\bf 16}, 748 (1966);
G.T. Zatsepin and V.A. Kuzmin, Zh. Eksp. Teor. Fiz. Pis'ma Red.
{\bf 4}, 414 (1966) [JETP Lett. {\bf 4}, 78 (1966)].
\bibitem{szabo} For a recent review on noncommutative QFT
see {\it e.g.} R. Szabo, hep-th/0109162.
\bibitem{nair} V.P. Nair and A.P. Polychronakos, Phys. Lett. B {\bf 505}, 
267 (2001).
\bibitem{hattfield} See {\it e.g.} B. Hatfield,
{\it Quantum Field Theory of Point Particles and Strings}, p. 199
(Benjamin, 1992).
\bibitem{connes} A. Connes, M.R. Douglas, A. Shwartz, J. High Energy
Phys. {\bf 02} (1999) 003.
\bibitem{sw} N. Seiberg, E. Witten, J. High Energy Phys. {\bf 09} (1999) 032.
\bibitem{mezincescu} L. Mezincescu, hep-th/0007046.
\bibitem{cg} S. Coleman and S.L. Glashow, Phys. Rev. D {\bf 59}, 116008
(1999); R. Aloisio, P. Blasi, P. Ghia and A. Grillo, Phys. Rev. D {\bf 62}, 
053010 (2000).
\bibitem{tri} For a review on the tritium beta-decay see {\it e.g.} 
J. Bonn and C. Weinheimer, Acta Phys. Polon. {\bf B31}, 1209 (2000).
\bibitem{abcc}
G. Amelino-Camelia and T. Piran, Phys. Rev. D {\bf 64}, 036005 (2001); 
O. Bertolami, Gen. Rel. Grav. {\bf 34}, 707 (2002);
J.M. Carmona and J.L. Cort\'es, Phys. Rev. D {\bf 65}, 025006 (2001).
\bibitem{halzen} For a recent review on cosmic rays see {\it e.g.} 
F. Halzen and D. Hooper, astro-ph/0204527.
\bibitem{kaon} T. Hambye, R. Man and U. Sarkar, Phys. Lett. B {\bf 421}, 
105 (1998).
\bibitem{kost} S. Carroll, J. Harvey, V. Kosteleck\'y, C. Lane and T. Okamoto, 
Phys. Rev. Lett. {\bf 87}, 141601 (2001); M. Chaichian, 
M. Sheikh-Jabbari and A. Tureanu, Phys. Rev. Lett. {\bf 86}, 2716 (2001);  
H. Falomir, J. Gamboa, M. Loewe, F. M\'endez and J.C. Rojas,
hep-th/0203260, to appear in Phys. Rev. D.
\bibitem{QTNCF} J.M. Carmona, J.L. Cort\'es, J. Gamboa and F. M\'endez,
J. High Energy Phys. {\bf 03} (2003) 058.
\end{thebibliography}
\end{document}